\documentstyle[amssymb,preprint,aps,epsf]{revtex}

\tighten
\draft

\begin{document}
\title{{\bf Wronskian Approach and the One-dimensional Schr\H{o}dinger Equation
with Double-well Potential}}
\author{Qiu Jian$^{1,a}$, Ru-Keng Su$^{2,1,b}$}
\address{$^{1}$Department of Physics, Fudan University, Shanghai 200433, P.R.China}
\address{$^{2}$China Center of Advanced Science and Technology (World aboratory),\\
P. O. Box 8730, Beijing, P.R.China}
\address{$^{a}$qiujian791102@hotmail.com}
\address{$^{b}$rksu@fudan.ac.cn}
\maketitle

\begin{abstract}
A Wronskian determinant approach is suggested to study the energy and the
wave function for one-dimensional Schr\H{o}dinger equation. An integral
equation and the corresponding Green's function are constructed. As an
example, we employed this approach to study the problem of double-well
potential with strong coupling. A series expansion of ground state energy up
to the second order approximation of iterative procedure is given.

keywords: double-well potential, Wronskian determinant
\end{abstract}


\section{\protect\bigskip {\bf Introduction}}

As a typical bound-state tunnelling problem, the solutions of
one-dimensional Schr\H{o}dinger equation with double-well potential are of
interest and have attracted much attention[1-9]. Recently, Friedberg, Lee
and Zhao(FLZ) suggested a new method[7-9] to decide the wave functions of
Schr\H{o}dinger equation by quadrature along a single trajectory. They gave
a good trial wave function for double-well potential and constructed the
Green's function. By means of the Green's function and the trial wave
function, they solved the integral equation by iterative procedure and found
the successive approximations for the true wave function and the
corresponding energy. They proved that the iterative solution is convergent
and gave the numerical result of the ground-state energy at first order
approximation of iterative procedure.

The objective of this paper is to address another method, namely, the
Wronskian determinant approach(WDA) to study this problem. We will give an
integral equation and the Green's function by a general argument. In a
specific case, our equation reduces to that equation given by FLZ method. We
will solve our integral equation by iterative procedure up to the second
order and give the numerical result of the energy. As a comparison, we will
use the trial wave function given by ref.[9] and the variational method to
calculate the ground state energy. The numerical results given by the three
methods(FLZ, WDA and variational method) are quite similar, the differences
occur in the higher orders only. This result confirms that the Wronskian
determinant approach is also successful for studying the solutions of
one-dimensional Schr\H{o}dinger equation.

\section{\bf Wronskian Determinant Approach}

Consider a second order ordinary differential equation: 
\begin{equation}
y^{^{\prime \prime }}+p(x)y^{^{\prime }}+q(x)y=f(x)\ \ \ \ \ x\epsilon [a,b]
\label{1}
\end{equation}
with the boundary conditions: 
\begin{equation}
\alpha y(a)+\beta y^{\prime }(a)=0\ \ \ \ \ \gamma y(b)+\delta y^{\prime
}(b)=0  \label{2}
\end{equation}
Assume $y_1$ is a solution of the corresponding homogeneous equation: 
\begin{equation}
y^{^{\prime \prime }}+p(x)y^{^{\prime }}+q(x)y=0  \label{3}
\end{equation}
One can easily prove that another linearly independent solution of Eq.(\ref
{3}) reads: 
\begin{equation}
y_2=y_1\int_a^xy_1^{-2}\Delta dx  \label{4}
\end{equation}
where $\Delta =\left( 
\begin{array}{ll}
y_1 & y_2 \\ 
y_1^{\prime } & y_2^{\prime }
\end{array}
\right) $ is the Wronskian determinant and it satisfies:

\begin{equation}
\Delta (x)=\Delta (a)\exp (-\int_a^xp(x)dx)  \label{5}
\end{equation}
Obviously, $y_2$ does not satisfy the boundary conditions, because it must
satisfy the uniqueness theorem. Suppose the solution $y$ of Eq.(\ref{1}) can
be written as: 
\begin{equation}
y=C_1(x)y_1+C_2(x)y_2  \label{6}
\end{equation}
where $C_1(x)$ and $C_2(x)$ are two unknown functions which satisfy a
constraint: 
\begin{equation}
C_1^{\prime }(x)y_1+C_2^{\prime }(x)y_2=0  \label{7}
\end{equation}
Since $y_2$ does not satisfy the boundary conditions, in order that $y$
satisfy Eq.(\ref{2}), we have: 
\begin{equation}
C_2(a)=C_2(b)=0  \label{8}
\end{equation}
Substituting Eq.(\ref{6}) and Eq.(\ref{7}) into Eq.(\ref{1}), noting that $%
y_1$ and $y_2$ are the solution of Eq.(\ref{3}), we find: 
\begin{equation}
C_1^{\prime }(x)y_1^{\prime }+C_2^{\prime }(x)y_2^{\prime }=f(x)  \label{9}
\end{equation}
Combining Eq.(\ref{7}) and Eq.(\ref{9}), we get: 
\begin{equation}
\left( 
\begin{array}{ll}
y_1 & y_2 \\ 
y_1^{\prime } & y_2^{\prime }
\end{array}
\right) \left( 
\begin{array}{l}
C_1^{\prime } \\ 
C_2^{\prime }
\end{array}
\right) =\left( 
\begin{array}{l}
0 \\ 
f(x)
\end{array}
\right)  \label{10}
\end{equation}
Hence: 
\begin{equation}
\left( 
\begin{array}{l}
C_1 \\ 
C_2
\end{array}
\right) =\int \left( 
\begin{array}{ll}
y_1 & y_2 \\ 
y_1^{\prime } & y_2^{\prime }
\end{array}
\right) ^{-1}\left( 
\begin{array}{l}
0 \\ 
f(x)
\end{array}
\right) dx  \label{11}
\end{equation}
\begin{equation}
C_1=-\int_B^x\frac 1\Delta y_2f(x)dx  \label{12}
\end{equation}
\begin{equation}
C_2=\int_A^x\frac 1\Delta y_1f(x)dx  \label{13}
\end{equation}
\begin{equation}
y=-y_1\int_B^x\frac 1\Delta y_2f(x)dx+y_2\int_A^x\frac 1\Delta y_1f(x)dx
\label{14}
\end{equation}
where $A$ and $B$ are two constants. From the identities Eq.(\ref{8}), we
have:

\begin{enumerate}
\item  The integraton bound A=a

\item  The integration: 
\begin{equation}
\int_a^b\frac 1\Delta y_1f(x)dx=0  \label{15}
\end{equation}
Eq.(\ref{15}) is called the 'resonant condition'.
\end{enumerate}

The integral bound 'B' is still undetermined. We will discuss this problem
in the next section.

\section{{\bf Schr\H{o}dinger equation}}

The one-dimensional Schr\H{o}dinger equation reads: 
\begin{equation}
-\frac 12\frac{d^2}{dx^2}\Phi +V\Phi =E\Phi  \label{16}
\end{equation}
The boundary conditions for bound-states are: 
\begin{equation}
\Phi (\infty )=\Phi (-\infty )=0  \label{17}
\end{equation}
Consider a perturbation equation: 
\begin{equation}
-\frac 12\frac{d^2}{dx^2}\Psi +(V+\omega )\Psi =(E+e)\Psi  \label{18}
\end{equation}
where $\omega $ is the perturbation potential and $e$ the energy shift. Now
we study how to find the bound-state wave function $\Psi $ and the energy $%
E+e$ provided $\Phi $ and $E$ are known. The boundary conditions for $\Psi $
reads: 
\begin{equation}
\Psi (\infty )=\Psi (-\infty )=0  \label{19}
\end{equation}
Rewrite Eq.(\ref{18}) as: 
\begin{equation}
-\frac 12\frac{d^2}{dx^2}\Psi +(V-E)\Psi =(e-\omega )\Psi  \label{20}
\end{equation}
and\ take the right hand side of Eq.(\ref{20}) as the inhomogeneous term. By
using Eq.(\ref{14}), we have: 
\begin{equation}
\Psi =-2\Phi \int_B^x\Phi _1(\omega -e)\Psi dx+2\Phi _1\int_{-\infty }^x\Phi
(\omega -e)\Psi dx  \label{21}
\end{equation}
where $\Phi _1=\Phi \int_0^x\Phi ^{-2}dx$, according to Eq.(\ref{4}) The
resonant condition now becomes: 
\begin{equation}
\int_{-\infty }^\infty \Phi (\omega -e)\Psi dx=0  \label{22}
\end{equation}

We can get the energy shift by the identity Eq.(\ref{22}) if $\Psi $ is
obtained. However, the lower bound B of the integral in Eq.(\ref{21}) is
undetermined. Notice that the different choices of B will only lead to an
extra constant times $\Phi $ in Eq.(\ref{21}). Then we can fix B=0 and
rewrite Eq.(\ref{21}) as: 
\begin{equation}
\Psi =c\Phi -2\Phi \int_0^x\Phi _1(\omega -e)\Psi dx+2\Phi _1\int_{-\infty
}^x\Phi (\omega -e)\Psi dx  \label{23}
\end{equation}

Combine the two integrations (using Eq.(\ref{22})): 
\begin{equation}
\Psi =c\Phi -2\Phi \int_0^x\Phi ^{-2}(y)dy\int_y^\infty \Phi (z)(\omega
-e)\Psi (z)dz  \label{24}
\end{equation}
There are two ways to determine the constant $c$:

\begin{enumerate}
\item[(i)]  Assume that the perturbation potential is zero at infinity, then 
$\frac{\Phi (\infty )}{\Psi (\infty )}=1,$and according to Eq.(\ref{24}): 
\begin{equation}
1=\frac{\Psi (\infty )}{\Phi (\infty )}=c-2\int_0^\infty \Phi
^{-2}(y)dy\int_y^\infty \Phi (z)(\omega -e)\Psi (z)dz  \label{25}
\end{equation}
Substitute Eq.(\ref{25}) back into Eq.(\ref{24}), we have: 
\begin{equation}
\Psi =\Phi -2\Phi \int_\infty ^x\Phi ^{-2}(y)dy\int_y^\infty \Phi (z)(\omega
-e)\Psi (z)dz  \label{26}
\end{equation}
This is the result given by FLZ method[9].

\item[(ii)]  The integral equation can be solved by iterative method, using $%
\Psi _0=\Phi $ as the zeroth approximation, the corresponding energy shift
is: 
\begin{equation}
e_1=\frac{\int_{-\infty }^\infty \Phi ^2\omega dx}{\int_{-\infty }^\infty
\Phi ^2dx}  \label{27}
\end{equation}
according to Eq.(\ref{22}). The first approximation of wave function $\Psi _1
$ is given by: 
\begin{equation}
\Psi _1=c_0\Phi -2\Phi \int_0^x\Phi ^{-2}(y)dy\int_y^\infty \Phi (z)(\omega
-e_1)\Phi (z)dz  \label{28}
\end{equation}
Put $c_0=1+N,$ where $N$ is chosen to be: 
\begin{equation}
N=\frac{2\int_{-\infty }^\infty \Phi ^2dx\int_0^x\Phi
^{-2}(y)dy\int_y^\infty \Phi (z)(\omega -e_1)\Phi (z)dz}{\int_{-\infty
}^\infty \Phi ^2dx}  \label{29}
\end{equation}
we choose $c_0$ to satisfy: 
\begin{equation}
\left\langle \Psi _1|\Phi \right\rangle =\left\langle \Phi |\Phi
\right\rangle   \label{30}
\end{equation}
which means the modification $(\Psi _1-\Phi )$ is orthorgonal to $\Phi $.
Similarly, $e_2$ is given by: 
\begin{equation}
e_2=\frac{\int_{-\infty }^\infty \Phi \Psi _1\omega dx}{\int_{-\infty
}^\infty \Phi \Psi _1dx}  \label{31}
\end{equation}
and the corresponding constant $c_1$is chosen to satisfy: 
\begin{equation}
\left\langle \Psi _2|\Phi \right\rangle =\left\langle \Phi |\Phi
\right\rangle   \label{32}
\end{equation}
and so on.This method of choosing $c$ features that $c$ is different in each
order of iterative procedure and the modification of each order is
orthorgonal to $\Phi $.
\end{enumerate}

\section{\bf Double-well Potential}

As an example, we will employ the wronskian determinant approach to discuss
the one-dimensional Schr\H{o}dinger equation with double-well potential.
Schr \H{o}dinger equation reads: 
\begin{equation}
-\frac 12\frac{d^2}{dx^2}\Psi +\frac 12g^2(x^2-1)^2\Psi =E\Psi  \label{33}
\end{equation}
where $g$ is a large parameter in the strong coupling case.

Since the potential is symmetric, we can solve it in the half space $%
x\geqslant 0$ only. So the boundary condition for the ground state becomes: 
\begin{equation}
\Psi (\infty )=\Psi ^{\prime }(0)=0  \label{34}
\end{equation}

The trial wave function given by ref.[9] is: 
\begin{equation}
\Phi =\frac 1{1+x}\exp (-\frac g3(x-1)^2(x+2))  \label{35}
\end{equation}

The derivative of $\Phi $ is not zero at the origin. We must do some
modification. Let: 
\begin{equation}
\tilde{\Phi}=\left\{ 
\begin{array}{c}
\Phi +\frac{g-1}{g+1}\exp (-\frac{4g}3)\exp (\frac g3(x-1)^2(x+2))\ for\
0<x<1 \\ 
\Phi +\frac{g-1}{g+1}\exp (-\frac{4g}3)\Phi \ \ \ \ \ \ \ \ \ \ \ \ \ \ \ \
\;\;\;\;\;for\ x>1
\end{array}
\right.  \label{36}
\end{equation}

The difference between $\tilde{\Phi}$ and $\Phi $ is of the order $\exp (-%
\frac{4g}3).$For strong coupling, $g\gg 1$, the difference is exponentially
small and can be neglected. We can still choose $\Phi $ as our trial
function. the wave function $\Phi $ satisfies: 
\begin{equation}
-\frac 12\frac{d^2}{dx^2}\Phi +(\frac 12g^2(x^2-1)^2+\frac 1{(1+x)^2})\Phi
=g\Phi   \label{37}
\end{equation}
Treat this equation as the unperturbed one, and $-\frac 1{(1+x)^2}$as the
perturbative potential $\omega .$ Employing the Wronsky determinant
approach, the integral equation reads: 
\begin{equation}
\Psi =c\Phi -2\Phi \int_0^x\Phi ^{-2}(y)dy\int_y^\infty \Phi (z)(\omega
-e)\Psi (z)dz  \label{38}
\end{equation}
The energy shift is given by: 
\begin{equation}
e=\frac{\int_0^\infty \Phi \Psi \omega dx}{\int_0^\infty \Phi \Psi dx}
\label{39}
\end{equation}
We solve the integral equation Eq.(\ref{38}) by iterative procedure.

\subsection{\bf First order approximation}

At the first order approximation, $\Psi _0=\Phi ,$the energy shift $e_1$ and
the $\Psi _1$ are: 
\begin{equation}
e_1=\frac{\int_0^\infty -\frac 1{(1+x)^4}\exp (-\frac{2g}3(x-1)^2(x+2))dx}{%
\int_0^\infty \frac 1{(1+x)^2}\exp (-\frac{2g}3(x-1)^2(x+2))dx}  \label{40}
\end{equation}
\begin{equation}
\Psi _1=(1+N)\Phi -2\Phi \int_1^x\Phi ^{-2}(y)dy\int_y^\infty \Phi
(z)(\omega -e)\Phi (z)dz  \label{41}
\end{equation}
where we have put the integration lower bound to be 1 instead of 0. The
difference can be cancelled by the modification of $N$. This alteration will
bring some convenience in the calculation. $N$ is also chosen to satisfy: 
\begin{equation}
\left\langle \Psi _1|\Phi \right\rangle =\left\langle \Phi |\Phi
\right\rangle   \label{42}
\end{equation}
For strong coupling, $g\gg 1$, using the integral formula: 
\begin{equation}
\int_{-a}^af(x)\exp (-gx^2)dx\symbol{126}f(0)\sqrt{\frac{2\pi }g}+\frac{%
f^{\prime \prime }(0)}4\sqrt{2\pi }g^{-3/2}+...+f^{(2n)}(0)\frac{\Gamma (%
\frac{2n+1}2)}{(2n)!}g^{-n-\frac 12}...  \label{43}
\end{equation}
we obtain a series of 1/g for energy: 
\[
e_1=-\frac 14-\frac 9{64}g^{-1}-0.1660g^{-2}-0.2755g^{-3}...
\]
and 
\[
E=g-\frac 14-\frac 9{64}g^{-1}-0.1660g^{-2}-0.2755g^{-3}...
\]
by a straight calculation.

\subsection{\bf Second order approximation}

At the second order approximation, the energy shift $e_2$ is: 
\begin{equation}
e_2=\frac{(1+N)\int_0^\infty \Phi \omega \Phi dx-2\int_0^\infty \Phi \omega
\Phi dx\int_1^x\Phi ^{-2}(y)dy\int_y^\infty \Phi (z)(\omega -e_1)\Phi (z)dz}{
\int_0^\infty \Phi ^2dx}  \label{44}
\end{equation}
We are facing on the triple integration in the numerator of Eq.(\ref{44}).
The detailed calculation of the triple integration is presented in the
appendix. The result is: 
\[
e_2=-\frac 14-\frac 9{64}g^{-1}-0.1738g^{-2}-0.3107g^{-3}... 
\]
and 
\[
E=g-\frac 14-\frac 9{64}g^{-1}-0.1738g^{-2}-0.3107g^{-3}... 
\]

\subsection{\bf Variational Method}

In this section, we use another method, namely, variational method, to
calculate the ground state energy. The trial function is chosen as: 
\begin{equation}
\Phi =\frac 1{1+x}\exp (-\frac f3(x-1)^2(x+2))  \label{45}
\end{equation}
where $f$ is the variational parameter. The ground state energy is: 
\begin{equation}
{\small E}=\frac{\int_0^\infty [\frac f{(1+x)^2}-\frac 1{(1+x)^4}+\frac 12%
(g^2-f^2)(x-1)^2]\exp (-\frac{2f}3(x-1)^2(x+2))dx}{\int_0^\infty \frac 1{%
(1+x)^2}\exp (-\frac{2f}3(x-1)^2(x+2))dx}  \label{46}
\end{equation}
The integration in both the numerator and denominator of Eq.(\ref{46}) can
be calculated by using Eq.(\ref{43}). The result is: 
\[
f=g-\frac 9{64}g^{-1}-0.3320g^{-2}-0.8366g^{-3}...
\]
and 
\[
E=g-\frac 14-\frac 9{64}g^{-1}-0.1660g^{-2}-0.2855g^{-3}-0.6184g^{-4}...
\]

\section{\bf Discussion and Conclusion}

In summary, we have studied the ground state energy and the wave function by
using the WDA. We have found that the Green's function and the integral
equation are the same as that of FLZ method provided the integral lower
bound B and parameter c are chosen as $\infty $ and 1 respectively. we
suggest a method to determine parameter c which warrant that the
modification of $\Phi $ is orthogonal to $\Phi $, which is similar to the
general perturbation method. We have expanded the energy as a series of $%
g^{-1}$ for double-well potential and obtained: 
\[
E=g-\frac{1}{4}-\frac{9}{64}g^{-1}-0.1660g^{-2}-0.2755g^{-3}...\ \ \ \ \
(1st\ order\ iterative\ approximation) 
\]
\[
E=g-\frac{1}{4}-\frac{9}{64}g^{-1}-0.1738g^{-2}-0.3107g^{-3}...\ \ \ \ (2nd\
order\ iterative\ approximation) 
\]
\[
E=g-\frac{1}{4}-\frac{9}{64}g^{-1}-0.1660g^{-2}-0.2855g^{-3}...\ \ \ \ \ \ \
\ \ \ \ \ \ \ \ \ \ (variational\ method) 
\]

The three results are quite similar to each other up to the order $O(g^{-1})$
. It is confirmed that the WDA is a successful method to investigate the
one-dimensional Schr\H{o}dinger equation and the alternative way of
determining the parameter c is reasonable. The differences of energy between
different methods and approximations occur at the order of $g^{-2}$ and
higher terms.

\section{\bf Appendix}

We will calculate the triple integration: 
\begin{equation}
\int_{0}^{\infty }\Phi \omega \Phi dx\int_{1}^{x}\Phi
^{-2}(y)dy\int_{y}^{\infty }\Phi (z)(\omega -e_{1})\Phi (z)dz  \label{47}
\end{equation}

Let $u:=x-1$, the integration becomes: 
\begin{eqnarray}
&&\int_{-1}^\infty \Phi \omega \Phi (u)du\int_0^u\Phi
^{-2}(v)dv\int_v^\infty \Phi (w)(\omega -e_1)\Phi (w)dw  \nonumber \\
&=&-\int_{-1}^\infty \frac 1{(2+u)^4}\exp
(-2g(u^2+u^3/3))du\int_0^u(2+v)^2\exp (2g(v^2+v^3/3))dv  \nonumber \\
&&\int_v^\infty \frac 1{(2+w)^2}(\omega -e_1)\exp (-2g(w^2+w^3/3))dw 
\nonumber \\
&=&\int_0^\infty ...\int_0^u...\int_v^\infty
...+\int_{-1}^0...\int_0^u...\int_v^\infty ...  \label{48}
\end{eqnarray}
which is the \bigskip Lagrange's inverse function theorem.

Suppose function $x=f(z)$ is analytic at the region: $|z-z_0|<\rho ,$-$\pi
+\delta <\arg (z-z_0)<\pi -\delta :$ 
\[
f(z)=f(z_0)+a_k(z-z_0)^k+a_{k+1}(z-z_0)^{k+1}+... 
\]
while $\varphi (z)$ is analytic in the neighbourhood of $z_0.$Then:

\begin{equation}
\varphi (z)=\varphi (z_0)+\sum\limits_1^n\frac 1{n!}\frac{d^{n-1}}{dx^{n-1}}
[\varphi ^{\prime }(z)(\frac{z-z_0}{(f(z)-f(z_0))^{1/k}}
)^n]|_{z=z_0}(x-x_0)^{n/k}  \label{50}
\end{equation}
\ where $x_0=f(z_0).$

When $\varphi (z)=z,$ Eq.(\ref{50}) gives the formulae of inverse function $%
z=f^{-1}(x).$ In our case, \bigskip Let: 
\[
x=u^2+u^3/3 
\]

We find: 
\begin{equation}
\frac 1{(u+2)^2}=1/4-\frac 14x^{1/2}+\frac{11}{48}x...  \label{51}
\end{equation}
\begin{equation}
\frac 1{(u+2)^4}=1/16-\frac 18x^{1/2}+\frac{17}{96}x...  \label{52}
\end{equation}
\begin{equation}
du=(\frac 12x^{-1/2}-1/6+\frac 5{48}x^{1/2}...)dx\ \   \label{53}
\end{equation}
Eqs.(\ref{51})\symbol{126}(\ref{53}) are only valid for $u>0$. When $u<0$,
they can be calculated similarly. Thus the first part of Eq.(\ref{48})
becomes: 
\begin{eqnarray}
&&\int_0^\infty (1/16-\frac 18x^{1/2})(\frac 12x^{-1/2}-1/6)\exp
(-2gx)dx\int_0^x(4+4y^{1/2})(\frac 12y^{-1/2}-1/6)  \nonumber \\
&&\exp (2gy)dy\int_y^\infty (1/4-\frac 14z^{1/2})(-1/4+\frac 14z^{1/2}-e_1)( 
\frac 12z^{-1/2}-1/6)\exp (-2gz)dz  \label{54}
\end{eqnarray}
The problem now is reduced to the calculating of the standard form: 
\begin{equation}
\int_0^\infty x^{m/2}\exp (-2gx)dx\int_0^xy^{n/2}\exp (2gy)dy\int_y^\infty
z^{p/2}\exp (-2gz)dz  \label{55}
\end{equation}
where m, n, p are integers ranging from -1 to infinity.

Rewrite Eq.(\ref{55}) as: 
\begin{eqnarray}
&&\int_0^\infty x^{m/2}\exp (-2gx)dx\int_0^xy^{n/2}\exp (2gy)dy\int_x^\infty
z^{p/2}\exp (-2gz)dz+  \nonumber \\
&&\int_0^\infty x^{m/2}\exp (-2gx)dx\int_0^xy^{n/2}\exp
(2gy)dy\int_y^xz^{p/2}\exp (-2gz)dz  \label{56}
\end{eqnarray}

In the first part, let: $x=zs,y=zst,z=z\ \ \ s\in [0,1]\ \ t\in [0,1]$ and\ $%
J=\left| \frac{\partial (x,y,z)}{\partial (z,s,t)}\right| =z^2s$

In the second part, let: $x=x,y=xst,z=xs\ \ $ $s\in [0,1]\ \ t\in [0,1]$
and\ $J=\left| \frac{\partial (x,y,z)}{\partial (x,s,t)}\right| =x^2s$ Eq.( 
\ref{56}) becomes: 
\begin{eqnarray}
&&\int_{D_1}s^{\frac{m+n+2}2}t^{n/2}z^{\frac{m+n+p+4}2}\exp \left[
-2gz\left( 1-st+s\right) \right] dzdsdt+  \label{57} \\
&&\int_{D_2}s^{\frac{m+p+2}2}t^{n/2}x^{\frac{m+n+p+4}2}\exp \left[
-2gx\left( 1-st+s\right) \right] dxdsdt  \nonumber
\end{eqnarray}
where 
\begin{eqnarray}
D_1 &:&\ z\in [0,\infty ),s\in [0,1],t\in [0,1] \\
D_2 &:&\ x\in [0,\infty ),s\in [0,1],t\in [0,1]
\end{eqnarray}
The variables x and z can be integrated Eq.(\ref{57}) becomes: 
\begin{eqnarray}
&&\frac{\Gamma (\frac{m+n+p+6}2)}{(2g)^{(m+n+p+6)/2}}\int s^{\frac{m+n+2}2
}t^{n/2}z^{\frac{m+n+p+4}2}\left( 1-st+s\right) ^{-\frac{m+n+p+6}2}dsdt+
\label{58} \\
&&\frac{\Gamma (\frac{m+n+p+6}2)}{(2g)^{(m+n+p+6)/2}}\int s^{\frac{m+p+2}2
}t^{n/2}x^{\frac{m+n+p+4}2}\left( 1-st+s\right) ^{-\frac{m+n+p+6}2}dsdt 
\nonumber
\end{eqnarray}

Now the parameter g is extracted out, the remaining two double integrations
are independent of g and the area for integration is a rectangle, they can
be calculated easily.

We calculate the integration which satisfy $\frac{m+n+p+6}2\leqslant k$ and
sum all these integrations, it will give us the first part of Eq.(\ref{48}).
The second part can be calculated with the same procedure.

We thank Prof. W.Q.Zhao and Prof. S.Z.Hu for helpful discussions. this work
was supported in part by NNSF of China under contracts No.19975050,
10047045, 19947001.

\end{document}